\documentclass[final,3p,times,authoryear]{elsarticle}

\usepackage{comment}
\usepackage{booktabs}
\usepackage{amssymb}
\usepackage{amsmath}
\usepackage{tipa}

\journal{Forensic Science International}

\begin{document}

\begin{frontmatter}

\title{Forensic deepfake audio detection using segmental speech features}

\author[inst1]{Tianle Yang\corref{cor1}}
\author[inst2]{Chengzhe Sun}
\author[inst2]{Siwei Lyu}
\author[inst3]{Phil Rose}

\affiliation[inst1]{
    organization={University at Buffalo},
    addressline={Department of Linguistics}, 
    city={Buffalo},
    postcode={14260},
    state={NY},
    country={United States}
}

\affiliation[inst2]{
    organization={University at Buffalo},
    addressline={Department of Computer Science and Engineering}, 
    city={Buffalo},
    postcode={14260}, 
    state={NY},
    country={United States}
}

\affiliation[inst3]{
    organization={Australian National University},
    addressline={Emeritus Faculty},
    city={Canberra},
    postcode={0200},
    state={ACT},
    country={Australia}
}

\begin{abstract}
This study explores the potential of using acoustic features of segmental speech sounds to detect deepfake audio. These features are highly interpretable because of their close relationship with human articulatory processes and are expected to be more difficult for deepfake models to replicate. The results demonstrate that certain segmental features commonly used in forensic voice comparison (FVC) are effective in identifying deep-fakes, whereas some global features provide little value. These findings underscore the need to approach audio deepfake detection using methods that are distinct from those employed in traditional FVC, and offer a new perspective on leveraging segmental features for this purpose. In addition, the present study proposes a speaker-specific framework for deepfake detection, which differs fundamentally from the speaker-independent systems that dominate current benchmarks. While speaker-independent frameworks aim at broad generalization, the speaker-specific approach offers advantages in forensic contexts where case-by-case interpretability and sensitivity to individual phonetic realization are essential.

\end{abstract}

\begin{keyword}
deepfake audio detection \sep deepfake speech \sep forensic voice comparison \sep likelihood ratio
\end{keyword}

\end{frontmatter}

\section{Introduction}
Rapid advancements in generative AI technologies, driven by deep learning, have brought significant benefits and serious concerns to society. Although generative AI enables applications such as voice cloning for virtual assistants, text-to-speech synthesis, and AI-driven customer support, it has also facilitated the proliferation of synthetic content, commonly known as {\it deepfakes}. These artificially generated audio, video, and image outputs are becoming increasingly realistic and have been exploited for malicious purposes, including impersonation and fraud. For example, \$240,000 was stolen by criminals using deepfake audio to impersonate a CEO of a UK company \citep{stupp2019fraudsters}. In another case, a high school teacher in Maryland created a fake audio clip of the principal making racist remarks, causing reputational harm and triggering public outrage \citep{finley2024deepfake}. Deepfakes have also been linked to political manipulation \citep{suwajanakorn2017synthesizing, ulmer2023deepfaking, kessler2020trump}, online harassment, and even undermining trust in evidence used in court or police investigations through fake audio \citep{rao2021review}.

To detect increasingly realistic deepfake audios, competitions such as ASVspoof \citep{wu2017asvspoof} and ADD \citep{yi2022add} were organized and benchmark datasets were created \citep{wu2015sas, reimao2019dataset, frank2021wavefake, zhao2024emofake}. Additionally, extensive efforts have been made to study features of acoustics that can aid detection models in identifying synthetic artifacts. Using these features, researchers have trained various detection models. Some rely on traditional classification algorithms, such as logistic regression \citep{rodriguez2020machine}, k-nearest neighbor \citep{singh2021detection}, and random forest \citep{ji2017ensemble}. Others use deep learning techniques, including convolutional neural networks \citep{hemavathi2021voice, wu2020light, sun2023ai}, deep residual networks \citep{alzantot2019deep}, and graph neural networks \citep{tak2021graph}.

According to a recent survey \citep{khanjani2023audio}, the acoustic features currently utilized for deepfake audio detection can be categorized into four groups: short-term spectral features (MFCC, LPS, LFCC, IMFCC), long-term spectral features (CQCC), prosodic features (F0, energy, duration), and self-supervised embedded features (XLS-R). These features are computationally efficient, making them well-suited for automated detection tasks. However, they primarily rely on abstract representations, which lack direct interpretability, making it difficult to understand the specific audio traits being analyzed, an issue particularly critical in forensic settings, where things often have to be explained to a judge or jury. In real forensic casework, the central question is often not simply whether a recording is synthetic, but whether it is a deepfake of a particular speaker. This frames the task as a speaker-dependent comparison rather than a population-wide screening problem. In what follows, we therefore adopt a speaker-specific design that models the same talker across sessions and conditions. This framing motivates the discussion that follows on transparency and fairness.

Because most deepfake audio detection models are trained based on these abstract features, they often operate as black-box systems that only provide a probability score for the audio being fake. This limits the transparency of the underlying evidence. In forensic science, especially in evidence evaluation, interpretable methods should be preferred because they support clear reasoning, reproducibility, and scrutiny. Legal standards such as the Daubert criteria \citep{farrell1993daubert}, together with guidance from ENFSI on evaluative reporting and international norms like ISO/IEC 30107 \citep{iso30107} on testing and disclosure, emphasize validation, transparency of procedures, and the ability to explain conclusions. Nonlinear models such as deep neural networks can be used when they are properly validated and documented, but their opacity can make courtroom communication and cross-examination more difficult.

By contrast, interpretable models, such as those based on likelihood ratios and well-understood acoustic or biometric features, allow experts to clearly articulate the basis of their assessments. Among these, phonetic features offer particular advantages due to their direct connection to physiological articulatory processes. For example, formant values (F1, F2, and F3) are closely linked to vowel articulation: F1 is inversely related to the height of the tongue, F2 is positively correlated with the frontness of the tongue, and F3 captures additional features, such as the rounding of the lips, which typically lowers its frequency, as well as other finer articulatory nuances \citep[pp. 347–350]{hardcastle2012handbook}. By integrating these linguistically grounded features, detection methods might provide transparent insights that are highly valuable in forensics. Such phonetic-based detection offers scientifically sound evidence to identify audio manipulation, ensuring accountability in legal contexts.

A parallel consideration is fairness. Recent studies document group-dependent performance differences for deepfake speech detectors, including bias by speakers' language, gender, age, and accent \citep{hutiri2022bias, yadav2024fairssd, moreno2025revealing, stanvek2025scdf}, and sensitivity to random dataset artifacts such as leading silence \citep{smeu2025circumventing}. Specifically, \cite{yadav2024fairssd} shows that commonly used synthetic-speech detectors display systematic demographic bias: higher false positive rates for certain genders, accents, age groups, and particularly for speakers with speech impairments. \cite{moreno2025revealing} extends this line of work to the cross-lingual setting, demonstrating that detectors trained only on English produce markedly uneven performance across ten languages. Spoofed speech in Romanian, Russian, French, and Finnish was far more likely to be detected as fake than English, German, or Swahili, even under identical synthesis conditions, revealing language identity as a latent bias factor in countermeasure systems. \cite{stanvek2025scdf}, introducing the SCDF dataset, further confirms that detection accuracy varies by speaker sex, age, and language across balanced samples of fifty speakers and five languages, underscoring that demographic variables directly influence deepfake detector behavior. Finally, \cite{smeu2025circumventing} exposes a dataset-level artifact in widely used multimodal deepfake benchmarks, showing that the mere presence of a brief leading silence in manipulated recordings allows a simple classifier to achieve over 98 percent accuracy, meaning many models may have learned to rely on such spurious cues rather than genuine speech characteristics.

In sum, because these systems are trained on large, imbalanced corpora with metric-learning objectives, they can inherit and amplify training-data bias, which leads to unequal error rates across speaker groups. These findings suggest that fairness needs to be evaluated and reported explicitly, with stratified analyses across demographic, linguistic, and channel conditions. Given these limitations, we argue that current deepfake audio detection models should not serve as primary forensic evidence in court, as their lack of transparency and demonstrated bias make them unreliable for decisive legal judgments.

In addition to the opaque versus interpretable and bias versus fairness issues mentioned about detection models, there are also reasons from the deepfake generation side that motivate us to use these segmental phonetic features. On the generation side, current state-of-the-art deepfake models are built based on text-to-speech (TTS), voice conversion (VC), or hybrids of the two \citep{wang2026asvspoof}. In all cases, the output acoustics are tightly constrained by the acoustic and lexical distributions of the training data. This dependence gives rise to several systematic weaknesses (discussed below) that are most clearly exposed by segment-anchored features.

First, a widely observed issue is that models mishandle accentual and dialectal realization. TTS systems often render high-frequency, in-domain words consistently, yet switch to inconsistent pronunciations or accent patterns for rare or out-of-domain items \citep{taylor2019analysis, he2022neural, zhou2024accented}. In VC, conventional mappings primarily transfer timbre and average prosody but do not reliably convert accentual targets. As a result, accent conversion is treated as a distinct, more challenging problem and requires explicit methods beyond standard VC \citep{aryal2014can, jin2023voice}. Additionally, perceptual studies on deepfake voices point to the same limitation, showing that prosodic cues and accent-specific features strongly shape how listeners judge synthetic speech \citep{bakkouche2025finding}. Such dialectal contrasts include, for example, the /u/-fronting characteristic of California English and the low-back (cot–caught) merger common in many Midwestern varieties, both of which are well documented in sociophonetics \citep{labov2006atlas, hall2011completion}. These errors manifest locally at the vowel segment level and are therefore detectable with phonetic measures tied to specific segments.

Second, the acoustics of an individual's voice are shaped by anatomical constraints such as vocal fold length and vocal tract geometry. Generative models trained to minimize average loss over many segment tokens tend to regress toward population means and to smooth idiosyncratic realizations. This “over-smoothing” tendency of spectrogram-predictive neural TTS is widely reported, implying a loss of fine phonetic detail and idiosyncrasy \citep{ren2022revisiting, kogel2023towards}. We speculate that this is why deepfakes can often sound generally human, yet sometimes fail to resemble a specific individual convincingly. In contrast, classic and contemporary sources link $f_0$ to vocal fold physiology and formant patterns to vocal-tract length and body size; these stable, speaker-specific configurations are precisely the kind of cues that segment-anchored measures can detect \citep{titze1989relation, zhang2021contribution, pisanski2014vocal}.

Third, socially conditioned phonetic variation is underrepresented and weakly controlled in deepfake models. Style choices such as creaky voice, breathy voice, or systematic phrase-final lengthening occur in context-dependent ways that reflect identity and register \citep{podesva2007phonation, yuasa2010creaky, gordon2001phonation}. For example, some speakers produce frequent phrase-final creak, or deploy localized creak near vowels at intonation phrase onsets \citep{redi2001variation}. Others maintain a stable breathy quality in a certain tone register \citep{rose2022modelling}. Furthermore, socioeconomic status has also been found to be a factor in individual phonetic realizations. A classic research would be \cite{labov1986social}, which shows people living in the same area of New York pronounce words differently due to their different socioeconomic status. Current generators do not encode these sociophonetic variables explicitly and instead rely on coarse latent style or prosody embeddings, which improves expressiveness but still lacks segment-conditioned, text-conditioned, or social-conditioned control \citep{zaidi2021daft, latif2021controlling, korotkova2024word}.

Taken together, these detection-side and generation-side limitations provide a principled motivation for the speaker-specific, segmental framework. By anchoring measurements to identifiable phonetic units, the analysis can expose accent drift, failure to reproduce anatomical baselines, and missing social-phonetic cues, while remaining transparent and reproducible in forensic settings.

Therefore, we present the first speaker-specific study that compares both phonetic and acoustic measures, including the midpoints of vowel formants (MF), long-term fundamental frequency (LTF0), long-term formant distribution (LTFD), and mel-frequency cepstral coefficients (MFCC) in terms of their function in deepfake detection. While these features are commonly used in FVC, our task is different: instead of speaker comparison, we classify real vs. synthetic speech. To evaluate the performance of our system, we compute the log-likelihood ratio cost (Cllr; \citealp{brummer2006application}) and the equal error rate (EER), with Cllr used here as a cost-based metric derived from model likelihoods, rather than speaker trial comparisons. We found that segmental features such as MF outperform global features such as LTFD, LTF0, and MFCC in detecting synthetic speech.\footnote{We use the term “global” to refer to MFCCs and other unsegmented or frame-based features, or distributional features such as LTF0, in contrast to segment-anchored features extracted with reference to specific phonetic segments (e.g., vowel formants or vowel durations). This terminology is consistent with conventions in forensic phonetics and related work \citep{chan2024long}.} These results suggest that a system based on interpretable phonetic features may offer high performance and transparency for real versus fake audio detection.

\section{Experimental setting}

\subsection{Real speech processing and alignment}
In addition to open-access datasets such as LJ Speech \citep{ljspeech17} and the M-AILABS Speech Dataset \citep{m-ailabs-dataset}, we collected multiple interview recordings of native US English speakers from various YouTube recording sessions to serve as our dataset. All YouTube speakers were anonymized using their name initials. The rationale for using two distinct types of data is to evaluate our method under both ideal and real-world conditions. Open-access datasets represent high-quality, studio-recorded speech, which approximates an ideal acoustic setting. In contrast, YouTube recordings often contain background noise, overlapping speech, and other imperfections that are more representative of conditions under which manipulated audio may be encountered in practice. While it is feasible for malicious actors to extract speech from publicly available recordings, it is far less likely that they would have access to clean, studio-quality recordings of their targets.

In addition to the ideal versus real-world motivation, we also use the two sets for other design reasons. The YouTube set provides multiple interviews of the same individual across different years and contexts, which allows us to examine within-speaker behavior over time under uncontrolled conditions. The open-access set, particularly LJ Speech, offers a single-speaker baseline that matches our speaker-specific framework and supports reproducibility, and M-AILABS has a similar format that keeps preprocessing and alignment consistent. We do not use telephony benchmarks commonly employed in forensic evaluations because they typically lack precise recordings of the same person across many years, which is essential for our design. Future studies can consider a more diverse set of recording conditions, including telephony speech datasets and recordings with ambient noise and compression artifacts.

As for segment-level alignment for the YouTube recordings, we first manually isolated each speaker’s speech from the recordings to ensure accuracy. In cases of overlapping speech, the entire overlapped portion was removed. Then, we segmented the isolated audio into smaller chunks to ensure the accuracy of transcriptions and alignments. We employed OpenAI's Whisper model (medium.en) \citep{radford2023robust} to generate initial transcriptions of the segmented audio and cross-checked them against the human captions available on YouTube. Then, we used the Montreal Forced Aligner (Montreal Forced Aligner; \citealp{mcauliffe2017montreal}) to align the audio with the transcribed text. Finally, we conducted a thorough manual review and quality control to ensure the accuracy of these transcriptions and alignments. Segment boundaries obtained from MFA were retained as-is, unless obvious alignment errors were detected during manual inspection. For quality control, any utterances with mismatched transcripts, corrupted audio, or segmentation issues were removed before feature extraction (8.9\% word tokens excluded) based on the cross-examination of the Whisper transcription with the YouTube official subtitles of the interview posted by the media. For the open-access datasets, as they are already segmented and transcribed, we directly applied the above MFA procedure on them.

\subsection{Deepfake audio generation}
We developed deepfake audio samples by training on real speech datasets and generating synthetic audio samples using ElevenLabs \citep{elevenlabs_2024_elevenlabs} and Parrot AI \citep{parrot}. For ElevenLabs, we employed their latest state-of-the-art speech synthesis model, Multilingual v2, which supports 29 languages and offers versatile capabilities including voice cloning, voice conversion, and text-to-speech. This model was selected for its ability to generate high-fidelity, multilingual audio with nuanced prosody and emotional expression. Similarly, we utilized Parrot AI's proprietary AI Voices generator, a state-of-the-art speech synthesis model leveraging a voice cloning and text-to-speech architecture. For both models, we trained on recordings from six real speakers (3 female, 3 male), generating over 250 diverse sample sentences per speaker. This resulted in a dataset of over 1,500 synthetic audio sentences, enabling a comparative study of twelve hypothetical speakers (6 real, 6 fake) against the original recordings.

\subsection{Acoustic extraction}
Audio samples were aligned by MFA into a TextGrid in Praat \citep{boersma2007praat}, and all the formant and $f_0$ data were extracted by parselmouth \citep{parselmouth}. The standard settings of the Praat Burg Formant Tracker were used (Maximum formant (Hz): 5000; Number of formants: 5; Window length (s): 0.025, Dynamic range (dB): 30; Pre-emphasis from (Hz): 50). First, the relevant $f_0$ and formant values (F1, F2, F3) were extracted from the vocalic portions of the audio. 

To ensure accuracy, we sampled $f_0$ and formant values at a sufficiently high frequency (15 equidistant points over duration), allowing us to capture fine temporal details of the data. By normalizing $f_{0}$ and formant trajectories in time rather than using fixed frame indices, we ensure that tokens with different vowel durations can be meaningfully compared on the same relative temporal scale\footnote{Following a widely adopted procedure in time-normalized speech analysis \citep[e.g.,][]{williams2014cross, yang2025onset}}.

The first 13 MFCCs (including the zeroth coefficient), alongside their corresponding delta and delta-delta coefficients (39 coefficients in total), were derived using the librosa Python library \citep{mcfee2015librosa}, with a 20 ms window length and 10 ms window shift with a frequency range from 0 to 8k Hz.

\subsection{Cllr and EER Calculation}
Following the approach of \cite{chan2024long}, we compute likelihood ratios (LRs) and EER, but instead of using a Gaussian Mixture Model–Universal Background Model (GMM-UBM), we adopt a simpler two-class GMM for each speaker. This is because GMM-UBM, though designed to model speaker variability via a shared background, is less suited to our binary classification task that contrasts real and synthetic speech. Since these two sources differ in origin and variability, we train separate GMMs for each class (e.g., real vs. fake formant) and compute the likelihood ratio for each token \( x \) as:

\begin{equation}
\text{LR}(x_t) = \frac{P(x_t \mid \text{real GMM})}{P(x_t \mid \text{fake GMM})}
\label{eq:lr}
\end{equation}

and define the log-likelihood ratio as:
\begin{equation}
\ell_t = \log \text{LR}(x_t) = \log \frac{P(x_t \mid \text{real})}{P(x_t \mid \text{fake})}
\label{eq:loglr}
\end{equation}

As the selection of number of Gaussians is empirical, we conducted a pre-test to choose the appropriate number, as suggested by \cite{chan2024long}. To evaluate detection performance, we calculate both the Cllr and the EER. Although Cllr is commonly used in FVC tasks involving same-speaker and different-speaker trials, we reinterpret it here as a cost-based metric that quantifies the discriminability between real and fake speech. In our setting, the goal is not to determine whether two recordings originate from the same speaker, but to classify whether a given speech token is real or synthetic. Accordingly, we designate real samples as “target” and fake samples as “non-target” trials purely as a labeling convention\footnote{The "target and non-target" here refer to bona fide and deepfake for our purpose}. Note that we assume flat prior odds for classifying each token as real or fake. The formula used to compute Cllr follows the standard definition \citep{brummer2006application, brummer2013bosaris}:

\begin{equation}
C_{\text{llr}} = 
\frac{1}{2|\mathcal{T}|} \sum_{t \in \mathcal{T}} \log_2\left(1 + e^{-\ell_t}\right) 
+ 
\frac{1}{2|\mathcal{N}|} \sum_{t \in \mathcal{N}} \log_2\left(1 + e^{\ell_t}\right)
\label{equation:Cllr}
\end{equation}

Here, \( \mathcal{T} \) and \( \mathcal{N} \) represent the sets of indices belonging to target and non-target trials, respectively. Unlike in FVC, where LRs are computed over speaker pairs, our scores derive directly from model likelihoods for individual speech tokens under two source hypotheses, $H_{\mathrm{real}}$ and $H_{\mathrm{fake}}$. We therefore use the canonical $\mathrm{Cllr}$ formula to quantify \emph{token-level evidential strength} for $H_{\mathrm{real}}$ versus $H_{\mathrm{fake}}$. 
This differs from speaker-comparison usage, but it is not merely a measure of decision uncertainty; it summarizes how well the scores for a given token support one source hypothesis over the other. 
We note that this is not a full LR system for court use, which would require additional adjustment, validation, and reporting steps.

\section{Overview of acoustic features}
This section presents an overview of the acoustic features examined in the study, using plots to illustrate patterns shared by and distinct between real and synthetic speakers. For clarity of presentation, the plots display a subset of vowels or tokens sampled from two illustrative speakers. In contrast, the formal analysis described later was based on data from twelve speakers (6 real, 6 fake). For each speaker, the full vowel inventory was analyzed, including both monophthongs and diphthongs (see Section 4 for details).

\subsection{Vowel formants}
\subsubsection{Formant midpoint}
\begin{figure}[t]
  \centering
  \includegraphics[width=\linewidth]{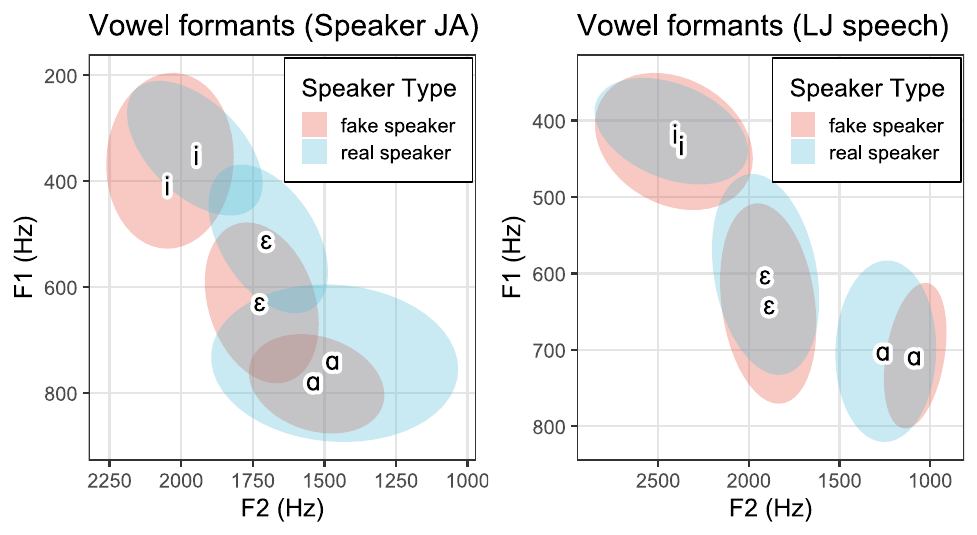}
  \caption{Ellipse plot (generated by the principal axes of the covariance matrices) of F1 (Hz) and F2 (Hz) values for selected vowel phonemes from two speakers' deepfake and real voice. The speaker JA (left) is from the YouTube group, and LJ Speech (right) is from the open-access datasets group.}
  \label{fig:vowel_mid}
\end{figure}

As we want to test our method with different recording settings and different speakers, we need to see how the vowel formants vary across these conditions. Figure \ref{fig:vowel_mid} illustrates this by showing the acoustics of three exemplar vowel phonemes from our speakers and their deepfakes. This chart came from data of mean formant values of 11,344 vowel tokens and was generated using the ellipse plot with a semi-transparent color representing a 75\% confidence region of the data points. These ellipses are meant to show the distribution and confidence areas of the vowel groups in the F2-F1 space. Specifically, the three vowels shown (/i, \textipa{E}, \textipa{A}/) typically account for about one-third to two-fifths of all vowel tokens; in our data, this corresponds to roughly 1.7k–2.3k tokens per speaker for these three vowels combined.

Figure \ref{fig:vowel_mid} clearly indicates that the formant values for the second speaker, LJ Speech, were captured more accurately compared to those for the first speaker, JA. It shows that the segmental quality of the deepfake output might vary depending on the recordings used to train the model.

Two factors may contribute to this difference. First, the background noise might play a role, as the LJ Speech dataset is recorded in a more acoustically controlled environment compared to the YouTube interview sessions. Additionally, the speakers' dialectal difference are expected to be a factor contributing to inaccuracies in the deepfake output, as they are likely not considered during the training of deepfake models.

\subsubsection{Long-term formant distribution}
\begin{figure}[t]
  \centering
  \includegraphics[width=\linewidth]{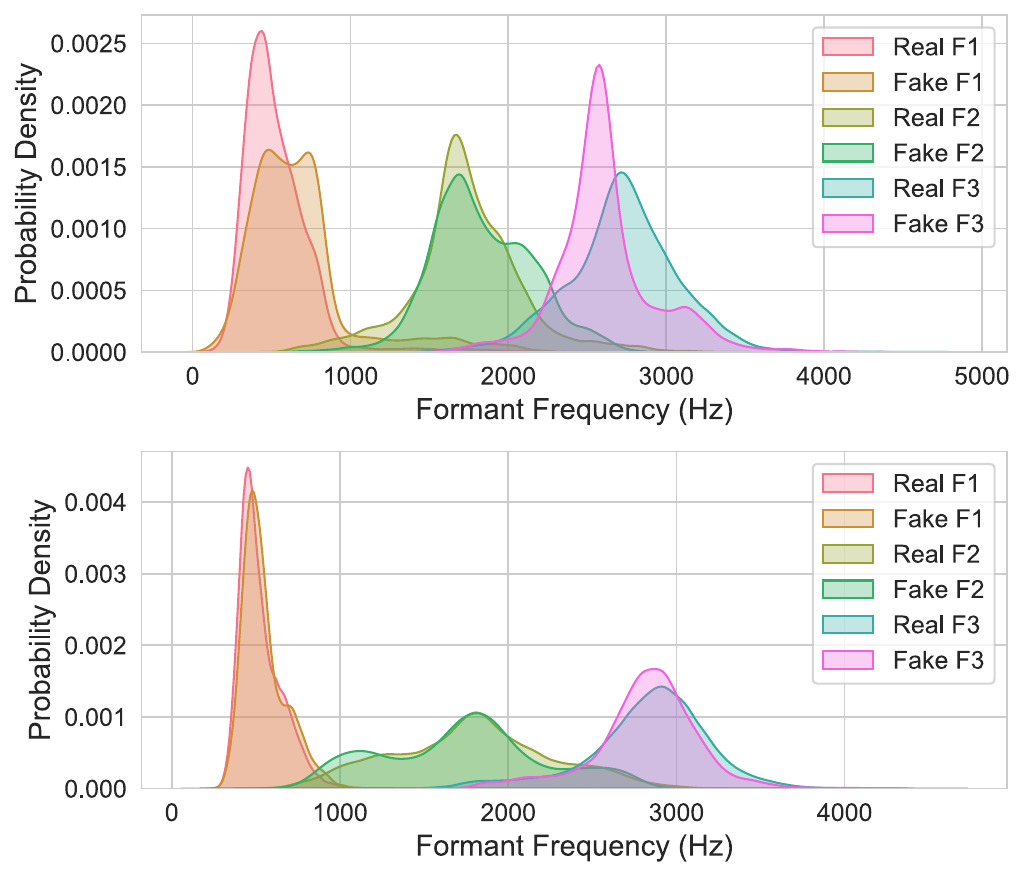}
  \caption{Long-term formant distributions of real and deepfake voices from two speakers: Speaker JA (top) from the YouTube group and LJ Speech (bottom) from the open-access datasets group.}
  \label{fig:LTFD}
\end{figure}

LTFD is a long-term phonetic feature that has been found to perform well in discriminating speakers \citep{french2015vocal, gold2013examining}. Figure \ref{fig:LTFD} illustrates such distributions, computed exclusively from the vocalic portions of speech rather than all phonetic segments. The figure shows that while some deepfakes (LJ speech, bottom panel) closely approximate the original speaker’s formant distribution, others (Speaker JA, upper panel) present more distinct deviations. A likely explanation is a mix of synthesis error and channel effects. If the cloning model was trained on noisy or compressed material, the learned spectral envelope can be biased toward those conditions; a mismatch between the training channel and the output channel can further shift long-term formant distributions. Codec artifacts, microphone response, and alignment inaccuracies on vowel segments may also contribute. We therefore treat Figure \ref{fig:LTFD} as illustrative of possible deviations; all performance claims are based on per-speaker quantitative metrics reported in the Section 4.

\subsection{Long-term fundamental frequency}
LTF0 is a commonly tested feature in FVC. However, its function in FVC is now criticized as the variability of f0 within a single speaker can be significantly influenced by various factors such as emotional states or health conditions \citep{braun1995fundamental}. Moreover, studies have shown that including LTF0 as a feature does not significantly enhance the strength of evidence in speech comparison \citep{kinoshita2005does, chan2024long}. In the context of deepfake detection, LTF0 is even less likely to be reliable, as the F0 of synthetic voices can be easily manipulated. Nonetheless, we chose to include this feature in our analysis to assess its utility.

\subsection{MFCC and Fbank}
MFCCs are sometimes used in FVC due to their ability to represent perceptually relevant spectral information. They are computed by applying a discrete cosine transform (DCT) to the log energies of a Mel-scaled filterbank (FBank), which distributes spectral resolution in a way that approximates human auditory sensitivity.

Although MFCCs are the features used in our experiments, they are not directly interpretable in terms of spectral energy distribution. To facilitate interpretation, we additionally present log-Mel FBank features as an illustrative tool. FBank retains the log energy of each filter prior to DCT, preserving the spectral shape and allowing clearer visual comparison of energy distribution across frequency bands.

\begin{figure}[t]
  \centering
  \includegraphics[width=\linewidth]{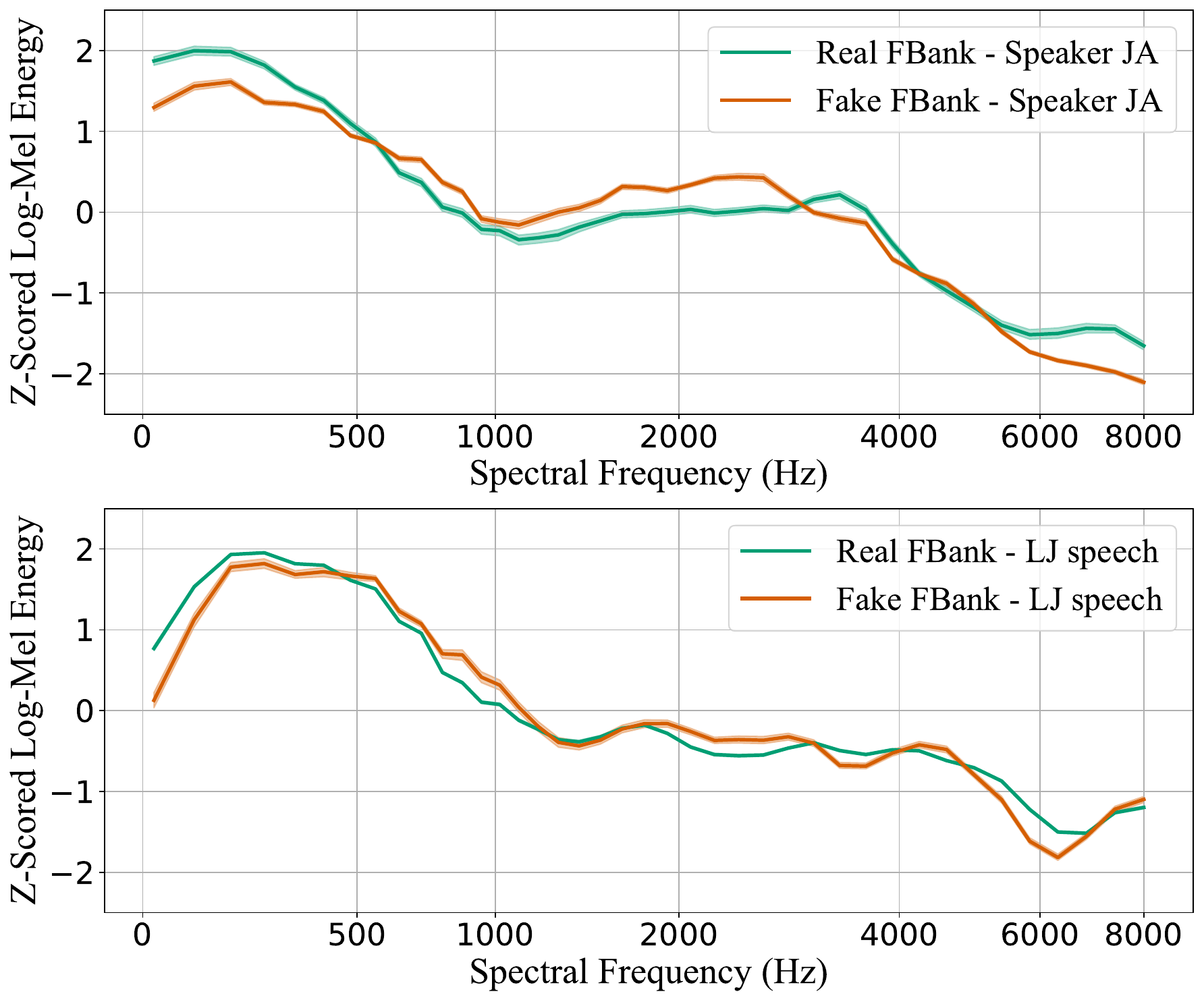}
  \caption{Z-scored log-Mel filterbank energy for real and fake utterances of two speakers (speaker JA - top panel, LJ speech - bottom panel). Shaded regions denote 95\% confidence intervals.}
  \label{fig:fbank}
\end{figure}

Figure \ref{fig:fbank} shows the z-scored log-Mel FBank energies\footnote{Note the x-axis represents the center frequencies of the mel filters converted into Hz for interpretability. Because the mel filters are evenly spaced in the perceptual mel scale but unevenly distributed in physical frequency (Hz), the spacing appears visually non-uniform, being denser at lower frequencies and sparser at higher ones. This is not a plotting error but a property of the mel-to-Hz mapping. } for two speakers, comparing real and synthetic utterances. The z-scoring was performed using the standard normalization formula, as illustrated in formula \ref{eq:zscore}, where $x_{i}$ denotes the raw log-Mel filterbank energy at band $i$, $\mu$ is the mean energy of that band, and $\sigma$ is the corresponding standard deviation.

\begin{equation}
z_{i} = \frac{x_{i} - \mu}{\sigma},
\label{eq:zscore}
\end{equation}

For each speaker, the differences between real and synthetic speech are minimal, and their spectral contours remain largely similar. In contrast, the differences between the two speakers are substantially more pronounced. These observations suggest that speaker identity has a stronger influence on FBank structure than the real versus synthetic distinction, and it is likely that real and fake speech of a speaker will have similar MFCCs. This is examined in detail in the next section.

\section{Cllr statistics for each feature}
In this section, we present the Cllr statistics for each feature. Since a person’s voice can change over time due to factors such as aging or health, we compared S1 vs. S2 samples (speech from the same speaker recorded in different sessions, and these sessions usually span more than 3 years) in addition to real vs. fake samples (authentic speech vs. synthesized or imposter speech). Utilizing non-contemporaneous recordings is vital in avoiding underestimating within-speaker variability, as within-speaker variation tends to increase over time.

The real vs. fake condition evaluates the system's ability to distinguish tokens from two distinct sources: real speech and deepfake speech. In this setup, we train two separate GMMs, one on real speech and one on fake speech\footnote{Note that GMMs used for LR computation in this study are trained separately for each speaker}. For each test token $x_1$, we compute a likelihood ratio as:

\begin{equation}
\text{LR}(x_1) = \frac{P(x_1 \mid \text{real GMM})}{P(x_1 \mid \text{fake GMM})}
\end{equation}

By contrast, the S1 versus S2 condition compares recordings from the same speaker in different sessions. Here, we train one GMM on the S1 data and another on the S2 data, and calculate likelihood ratios between them:

\begin{equation}
\text{LR}(x_2) = \frac{P(x_2 \mid \text{S1 GMM})}{P(x_2 \mid \text{S2 GMM})}
\end{equation}

Note that the real GMM = S1 GMM, and the S1 vs. S2 serves as a baseline condition here. Because the S1 and S2 recordings originate from the same speaker, they may be more acoustically similar. In contrast, real and fake speech may differ more substantially if the deepfake model fails to capture key acoustic features. Under such conditions, the system might be more uncertain in the S1 vs. S2 condition because the acoustic spaces of the two real speaker models overlap, leading to a higher $C_{\text{llr}}$ compared to real vs. fake group.

Moreover, in order to interpret the Cllr value and its relation with evidential strength, we need to define what constitutes a 'good' Cllr value first. According to \cite{van2024overview}, the usage of Cllr heavily depends on the field of forensics. For speech recognition datasets, the best Cllr is usually below 0.4 for strong evidential strength, while for signature comparison, the value is around 0.6. Furthermore, this value is empirical as it varies by algorithm (e.g., number of Gaussians) and sample size. For our purpose, we define a Cllr below 0.4 as good, between 0.4 and 0.6 as moderate, and above 0.6 as weak.

From Table \ref{tab:Cllr_eer_stats_1}, we observe that the voices of YouTube speakers exhibit temporal variation, with speakers JA and AM showing more rapid changes. Across all speakers, the phoneme [\textipa{U}] consistently yields the lowest Cllr values, while the second-lowest phoneme varies between individuals, for example, [i] for JA and [\textipa{aU}] for OB. A possible explanation is that [\textipa{U}] occupies an interior region of the vowel space that is sensitive to coarticulation, stress, and rounding dynamics. Small shifts in F1-F2 caused by smoothing in the synthesis pipeline, codec effects, or alignment noise can further push synthetic [\textipa{U}] away from the natural tokens, which improves separability and lowers Cllr. Training imbalance may also play a role if clean, sustained [\textipa{U}] tokens are underrepresented relative to highly frequent corner vowels. We note that this pattern does not imply that [\textipa{U}] is intrinsically more speaker specific. Rather, it suggests a mismatch that current systems have not yet captured for this category. Future work could test this hypothesis with controlled carrier phrases, balanced stress, and cross-system or cross-model evaluations, and could also repeat the analysis under band-limited and noisy conditions to probe robustness.

\begin{table*}[ht]
\small
\caption{Descriptive statistics of Cllr (with SD) and EER values for top 2 vowels with lowest Cllr, LTFD, LTF0, and MFCCs across 30 repetitions for four YouTube speakers.}
\label{tab:Cllr_eer_stats_1}
\centering

\begin{minipage}[t]{0.48\textwidth}
\centering
\textbf{Subject: JA (S1 vs. S2)} \\[2pt]
\begin{tabular}{lccc}
\toprule
\textbf{Feature} & \textbf{Cllr (mean)} & \textbf{SD} & \textbf{EER (\% mean)} \\
\midrule
MF [\textipa{U}]  & 0.512  & 0.1308   & 16.8    \\
MF [i]           & 0.661  & 0.0593   & 20.8    \\
LTFDs            & 0.804  & 0.0014   & 28.3    \\
LTF0             & 0.908  & 0.0004   & 37.4    \\
MFCCs             & 0.911 & 0.1402  &  37.2    \\
\bottomrule
\end{tabular}

\vspace{0.5em}
\textbf{Subject: JA (real vs. fake)} \\[2pt]
\begin{tabular}{lccc}
\toprule
\textbf{Feature} & \textbf{Cllr (mean)} & \textbf{SD} & \textbf{EER (\% mean)} \\
\midrule
MF [\textipa{U}]  & 0.258  & 0.0863   & 5.8     \\
MF [i]           & 0.404  & 0.0265   & 10.5    \\
LTFDs            & 0.661  & 0.0005   & 21.1    \\
LTF0             & 0.898  & 0.0005   & 38.1    \\
MFCCs             & 0.761 & 0.1251  &  27.5    \\
\bottomrule
\toprule
\end{tabular}
\end{minipage}
\hfill
\begin{minipage}[t]{0.48\textwidth}
\centering
\textbf{Subject: OB (S1 vs. S2)} \\[2pt]
\begin{tabular}{lccc}
\toprule
\textbf{Feature} & \textbf{Cllr (mean)} & \textbf{SD} & \textbf{EER (\% mean)} \\
\midrule
MF [\textipa{U}]  & 0.662  & 0.0587   & 21.2    \\
MF [\textipa{oU}] & 0.819  & 0.0205   & 30.0    \\
LTFDs            & 0.937  & 0.0002   & 37.8    \\
LTF0             & 0.946  & 0.0003   & 37.7    \\
MFCCs             & 0.924 & 0.1162  &  38.0    \\
\bottomrule
\end{tabular}

\vspace{0.5em}
\textbf{Subject: OB (real vs. fake)} \\[2pt]
\begin{tabular}{lccc}
\toprule
\textbf{Feature} & \textbf{Cllr (mean)} & \textbf{SD} & \textbf{EER (\% mean)} \\
\midrule
MF [\textipa{U}]  & 0.302  & 0.0393   & 7.3     \\
MF [\textipa{aU}] & 0.614  & 0.0272   & 18.1    \\
LTFDs            & 0.871  & 0.0005   & 32.6    \\
LTF0             & 0.859  & 0.0002   & 31.4    \\
MFCCs             & 0.889 & 0.0902  &  35.1    \\
\bottomrule
\toprule
\end{tabular}
\end{minipage}

\vspace{1em}

\begin{minipage}[t]{0.48\textwidth}
\centering
\textbf{Subject: AM (S1 vs. S2)} \\[2pt]
\begin{tabular}{lccc}
\toprule

\textbf{Feature} & \textbf{Cllr (mean)} & \textbf{SD} & \textbf{EER (\% mean)} \\
\midrule
MF [\textipa{U}]  & 0.558  & 0.1019   & 17.6    \\
MF [\textipa{3}]           & 0.623  & 0.0912   & 19.1    \\
LTFDs            & 0.772  & 0.0009   & 25.1    \\
LTF0             & 0.918  & 0.0005   & 36.9    \\
MFCCs             & 0.905 & 0.1326  &  35.8    \\
\bottomrule
\end{tabular}

\vspace{0.5em}
\textbf{Subject: AM (real vs. fake)} \\[2pt]
\begin{tabular}{lccc}
\toprule
\textbf{Feature} & \textbf{Cllr (mean)} & \textbf{SD} & \textbf{EER (\% mean)} \\
\midrule
MF [\textipa{U}]  & 0.214  & 0.0776   & 4.4     \\
MF [\textipa{aU}]   & 0.573  & 0.1435   & 16.8     \\
LTFDs            & 0.705  & 0.0013   & 21.8     \\
LTF0             & 0.911  & 0.0002   & 36.1     \\
MFCCs             & 0.870 & 0.1627  &  31.5    \\
\bottomrule
\end{tabular}
\end{minipage}
\hfill
\begin{minipage}[t]{0.48\textwidth}
\centering
\textbf{Subject: NH (S1 vs. S2)} \\[2pt]
\begin{tabular}{lccc}
\toprule
\textbf{Feature} & \textbf{Cllr (mean)} & \textbf{SD} & \textbf{EER (\% mean)} \\
\midrule
MF [\textipa{U}]  & 0.714  & 0.0915   & 28.3     \\
MF [\textipa{oU}] & 0.770  & 0.1697   & 29.0     \\
LTFDs            & 0.929  & 0.0005   & 37.4     \\
LTF0             & 0.955  & 0.0003   & 39.0     \\
MFCCs             & 0.902 & 0.0851  & 34.9    \\
\bottomrule
\end{tabular}

\vspace{0.5em}
\textbf{Subject: NH (real vs. fake)} \\[2pt]
\begin{tabular}{lccc}
\toprule
\textbf{Feature} & \textbf{Cllr (mean)} & \textbf{SD} & \textbf{EER (\% mean)} \\
\midrule
MF [\textipa{U}]  & 0.311  & 0.0751   & 7.5     \\
MF [\textipa{0}] & 0.572  & 0.0553   & 17.4     \\
LTFDs            & 0.841  & 0.0005   & 30.6     \\
LTF0             & 0.913  & 0.0003   & 37.6     \\
MFCCs             & 0.869 & 0.1209  &  30.9    \\
\bottomrule
\end{tabular}
\end{minipage}

\end{table*}

\begin{table*}[ht]
\small
\caption{Descriptive statistics of Cllr (with SD) and EER values for top 2 vowels with lowest Cllr, LTFD, LTF0, and MFCCs across 30 repetitions for two open-access dataset speakers.}
\label{tab:Cllr_eer_stats_2}
\centering

\begin{minipage}[t]{0.48\textwidth}
\centering
\begin{tabular}{@{}l@{}}
\textbf{Subject: LJ Speech (S1 vs. S2)} \\
\end{tabular}

\vspace{2pt}
\begin{tabular}{lccc}
\toprule
\textbf{Feature} & \textbf{Cllr (mean)} & \textbf{SD} & \textbf{EER (\% mean)} \\
\midrule
MF [\textipa{U}]  & 0.701  &  0.0257    &    22.2    \\
MF [\textipa{5}]  &  0.918  &  0.0075     &   35.9    \\
LTFDs  &  0.972    &   0.0002    &    42.0   \\
LTF0   &  0.984  &  0.0003   &   45.5    \\
MFCCs             & 0.947 & 0.0645  &  44.3    \\
\bottomrule
\end{tabular}

\vspace{0.5em}
\begin{tabular}{@{}l@{}}
\textbf{Subject: LJ Speech (real vs. fake)} \\
\end{tabular}

\vspace{2pt}
\begin{tabular}{lccc}
\toprule
\textbf{Feature} & \textbf{Cllr (mean)} & \textbf{SD} & \textbf{EER (\% mean)} \\
\midrule
MF [\textipa{U}] &  0.420   &  0.0650   &  11.1  \\
MF [\textipa{6}]   & 0.762   &  0.0458   &  27.8  \\
LTFDs         &  0.936  &   0.0004    &  37.6  \\
LTF0            &  0.961  & 0.0003   & 42.0  \\
MFCCs             & 0.923 & 0.0918  &  36.7    \\
\bottomrule
\end{tabular}
\end{minipage}
\hfill
\begin{minipage}[t]{0.48\textwidth}
\centering
\begin{tabular}{@{}l@{}}
\textbf{Subject: M-AILABS Speech (S1 vs. S2)} \\
\end{tabular}

\vspace{2pt}
\begin{tabular}{lccc}
\toprule
\textbf{Feature} & \textbf{Cllr (mean)} & \textbf{SD} & \textbf{EER (\% mean)} \\
\midrule
MF [\textipa{U}]   &  0.713  &  0.0675  & 23.1   \\
MF [\textipa{0}]   & 0.834  &  0.0519   &  30.1  \\
LTFDs   & 0.983  &  0.0003   & 43.9   \\
LTF0    & 0.970  &  0.0003   &  42.5  \\
MFCCs             & 0.951 & 0.0731  &  48.9    \\
\bottomrule
\end{tabular}

\vspace{0.5em}
\begin{tabular}{@{}l@{}}
\textbf{Subject: M-AILABS Speech (real vs. fake)} \\
\end{tabular}

\vspace{2pt}
\begin{tabular}{lccc}
\toprule
\textbf{Feature} & \textbf{Cllr (mean)} & \textbf{SD} & \textbf{EER (\% mean)} \\
\midrule
MF [\textipa{U}]   & 0.437  &  0.0332   & 15.1   \\
MF [\textipa{aU}]   & 0.635  &  0.0259   &  19.7  \\
LTFDs   & 0.809  &  0.0004   &  28.2  \\
LTF0    & 0.966  &  0.0004   &  42.1  \\
MFCCs             & 0.897 & 0.1172  &  34.2    \\
\bottomrule
\end{tabular}
\end{minipage}

\end{table*}

Hypothetically, if a deepfake model could accurately replicate natural phonetic features, we would expect the Cllr for real vs. fake speech to be higher than for S1 vs. S2, because the fake speech is trained on S1, and S1 vs. S2 group may contain greater acoustic variation over time. However, our results show the opposite pattern. For the MF feature [\textipa{U}], the Cllr in the real vs. fake condition is about half that in the S1 vs. S2 condition across four speakers. All other features listed show the same trend: lower Cllr for real vs. fake than for S1 vs. S2. In the S1 vs. S2 condition, both models are trained on real speech from the same speaker. If their acoustic properties are similar, the LRs from the two models will be close. This makes it harder for the system to make a confident choice, resulting in higher Cllr values. In contrast, in the real vs. fake condition, the two models represent distinct speech types. Their likelihoods are often farther apart, making it easier for the system to assign strong LRs, which leads to lower Cllr values. This suggests that the deepfake model does not fully capture some phonetic details.

In Table \ref{tab:Cllr_eer_stats_2}, we observe similar patterns indicating that MF features provide good evidential strength for deepfake detection. The major difference between Table \ref{tab:Cllr_eer_stats_1} and Table \ref{tab:Cllr_eer_stats_2} is their Cllr and EER ranges. The open-access dataset speaker (OA group) exhibits generally higher Cllr and EER values compared to our YouTube speakers (YouTube group). This outcome is likely due to the fact that the S1 vs. S2 samples in the OA dataset originate from similar recording sessions with limited temporal variation, which makes it harder for the system to distinguish between different conditions and results in a higher Cllr. Additionally, as open-access datasets typically consist of clean, professionally recorded speech collected for speech recognition or synthesis training, they tend to be more challenging for detection systems, leading to higher Cllr values for the real vs. fake comparison as well. Nevertheless, the MF feature [\textipa{U}] still shows approximately 40\% lower Cllr in the real vs. fake comparison relative to the S1 vs. S2 condition, suggesting that it retains considerable discriminative power even in this more challenging scenario.

For the global features like LTFDs, LTF0, and MFCCs, their evidential strength varies by speaker. For example, we observe better evidential strength for speaker JA’s LTFDs relative to the other speakers (see Table \ref{tab:Cllr_eer_stats_1}), but this improvement is neither universal nor large. For LTF0, it performs even worse compared to the LTFDs, with a Cllr higher than 0.85 for the YouTube group and above 0.95 for the OA group. For MFCCs, its evidential strength is between LTF0 and LTFDs, which also can be regarded as very weak.

In sum, all the long-term features tested in this study can be viewed as having weak evidential strength for deepfake detection, particularly for the OA group, where their LTFDs, LTF0, and MFCCs are almost identical.

\section{Discussion}

The findings of this study underscore the potential of using segmental acoustic features for enhancing the forensic detection of deepfake audio. The results demonstrate that phonetic features, such as vowel formants can provide good evidential value in distinguishing between genuine and synthetic speech. One key insight is the pronounced variability in the accuracy of deepfake models when replicating vowel formants. Based on this finding, we speculate that although there is a decent chance that the deepfake models can mimic most phones to some extent, the dialectical difference and the individual difference on the segmental level cannot be accurately captured by the deepfake models.

For example, we have seen that the deepfakes of LJ speech almost perfectly captured the speaker's LTFDs (Figure \ref{fig:LTFD}), but did not capture some MF accurately. In figure \ref{fig:vowel_mid}, LJ speech (real) has a higher F2 than the fake one when pronouncing the vowel [\textipa{A}], which means that the real speaker's tongue position is more front than the fake one (assuming the fake speaker has a tongue) when pronouncing [\textipa{A}]. This articulatory level difference is something that a model cannot easily pick up, as they likely were not considered during the model training process. This finding is particularly significant for forensic applications, where interpretability and transparency are crucial. Unlike black-box detection models relying on abstract features, segmental features offer interpretable evidence that aligns closely with human articulatory processes, providing a scientifically grounded basis for forensic analysis.

This method is theoretically supported by the fact that individuals have unique ways of realizing their phoneme inventories, influenced by various factors. These factors include: (1) physical differences in articulatory organs, such as male or taller speakers generally exhibiting lower formant frequencies; (2) dialectal variation, as seen in U.S. English, where California English speakers often exhibit fronting of the vowel /u/ (e.g., in dude), while Midwestern speakers commonly merge /\textipa{A}/ and /\textipa{O}/, making cot and caught homophones; and (3) individual variability. Given these influences, we analyzed the Cllr on a per-speaker basis, as these features vary significantly between speakers and are poorly replicated by deepfake models.

Although previous research has shown global features such as LTFDs or MFCCs to have good performance in real speaker comparisons \citep{french2015vocal, gold2013examining, chan2024long}, these features exhibited limited evidential strength in deepfake detection. The high Cllr values for these features reinforces the notion that long-term global features alone are insufficient for deepfake detection and should be supplemented by linguistically informed segmental measures. For other global features such as linear prediction cepstral coefficients (lpccs), we also expect them to have limited usages in deepfake detection as they process a speech signal by a certain short window length and window shift, which is commonly used for training speech production models. However, more studies are needed to assert their function.

Segmental features also have some limitations. They are time-consuming to compute and review, which reduces scalability for high-volume or urgent cases and makes them unsuitable for real-time detection. Our intended use is forensic analysis rather than large-scale screening. The approach further depends on accurate transcripts and phone alignments, which are the steps that cost the most time in our study. This dependence raises generalizability concerns outside U.S. English, since the current tools we use, such as MFA and Whisper, perform worse for underrepresented languages and dialects, and human quality control would need to be more extensive in those settings. Noisy or bandwidth-limited audio makes high-quality alignments harder to obtain, which can lower feature reliability. Future work could extend the protocol to additional languages with language-appropriate resources and native-speaker verification, and to test more alignment-light features such as long-term distributions that require minimal segmentation.

In addition, this framework also relies on obtaining enough vowel tokens per speaker to support stable estimation. Although the present study did not fix a hard threshold, our experience suggests that performance becomes more stable once the questioned recording yields several hundred usable vowel tokens. As a conservative guideline for casework, this typically corresponds to approximately ten minutes of continuous speech for the questioned (putatively synthetic) material, with a comparable amount of verified real speech from the same individual to fit the reference distributions. The exact requirement will vary with recording quality, channel, and speaking style. Future work should systematically determine token and duration thresholds under controlled conditions, and explore procedures that preserve evidential strength when only shorter recordings are available.

In order to get a stronger evidential strength, several acoustic features can be combined together to get a lower Cllr value, this step is called "feature fusion" \citep{brummer2013bosaris}. We did not implement this method here as the long-term features provided little evidential strength. However, more acoustic features could be tested in future research and combined together to acquire a system that has a better performance.

In summary, this study underscores a key finding: global features tested here show substantially weaker evidential strength than segmental features in terms of deepfake detection. For future acoustic studies on deepfake speech, incorporating additional cues such as speaker-specific intonation, tone, or pause patterns may enhance discriminative performance.

\section{Limitation}
A natural question concerns how comprehensive the results are, given that the proposed framework is speaker-specific. The answer is that the scope of generalization derives from the procedure rather than from a single pooled detector. The study advances a consistent forensic framework that can be applied to any suspected deepfake case, regardless of the particular synthesis method or dataset. The unit of analysis is the individual case, and the same protocol can be executed repeatedly across cases.

The approach is framework-driven rather than dataset-driven. For each speaker, we fit separate models to real and synthetic segmental distributions using a fixed feature set, a fixed likelihood-ratio logic, and a fixed evaluation procedure. This guarantees interpretability and reproducibility on a case-by-case basis. Because the framework is speaker-specific, it does not yield a single speaker-independent classifier and is therefore not directly comparable to benchmarks such as ASVspoof or ADD, which assume speaker-independent generalization at the population level. This is not a deficiency of the method but a consequence of the forensic inference target. In forensic practice, generalization means that a consistent and scientifically motivated procedure can be replicated across different speakers and cases, not that a single model generalizes across the population. Within this design, our experiments show that the same procedure reliably separates real and synthetic speech for each speaker examined. The claim is not that a population-level error rate has been established under benchmark assumptions, but that the framework is applicable and reliable on individual cases under a transparent and interpretable protocol.

Future work can extend coverage by executing the same protocol over larger speaker pools, more synthesis tools, and more recording conditions, and by packaging such evaluations into a forensic-style benchmark that permits community comparison while preserving interpretability. Elements of the framework may also be adapted to speaker-independent settings, provided that transparency and casewise auditability are retained.

\section{Conclusion}
In this study, we show that segmental features have sufficient evidence strength for forensic deepfake audio detection. These segmental features can't be easily captured by deepfake models compared to global acoustic features tested in this study. By incorporating these features, future methods can achieve higher transparency, interpretability, and reliability, ultimately enhancing the forensic community’s ability to address the growing challenge of deepfake audio, even in the most challenging forensic scenarios.

\newpage

\appendix
\section{Software setup}
\label{app1}

All the speech models and toolkits used in this study are publicly available online, and we have provided relevant references in the experimental setting section for the purpose of reproduction.


\bibliographystyle{elsarticle-harv}
\bibliography{references}

\end{document}